\documentclass[aip,pre,reprint,groupedaddress]{revtex4-1}

\usepackage{amsmath}    
\usepackage{amssymb}    
\usepackage{graphicx} 

\begin{document}

\title{Pinned Brownian Bridges in the Continuous-Time Limit}

\author{P.J. Malsom} 
\altaffiliation[Current address: ]{Kalamazoo Valley Community College, 6767 W O Ave, Kalamazoo, MI 49009}

\author{F.J. Pinski}
\email{frank.pinski@uc.edu}
\affiliation{Department of Physics, University of Cincinnati, Cincinnati, Ohio 45221, USA}
\date \today
 
\begin{abstract}

The current understanding of pinned Brownian bridges is based on the Onsager-Machlup (OM) functional.
The continuous-time limit of the OM functional can be expressed either by using the Fokker-Planck equation or by using the Radon-Nikodym derivative with the help of the Girsanov theorem and Ito's lemma.
The resulting expression, called here, the Ito-Girsanov (IG) expression, has been used as a basis of algorithms designed to create ensembles of transition paths, paths that are constrained to start in one free energy basin and end in another.
Here we explore the underlying formalism and show that the IG expression originates in a limit that is only conditionally convergent.
Thus without a sound mathematical foundation,  the IG expression produces unphysical results when used in computer algorithms that are designed to elucidate chemical transitions.
The consequence of this underlying mathematics is that the flaws in computer algorithms designed around the IG expression cannot be removed.
Despite the similarity between the theory of these pinned classical paths and quantum paths,  the continuous-time limits of each differ.
In particular, the factor of $i=\sqrt{-1}$ in the quantum action removes the infinities that are present when considering classical Brownian paths, removing the conditions on the convergence of continuous-time limit of the quantum propagator.
\end{abstract}


\maketitle


As a function of some external parameter, such as the temperature, the equilibrium state of a molecule may change, thereby modifying its physical properties.
In such events, the energy barrier between the starting and ending free energy basins limits the transition rate.
When the thermal energy is small compared to the barrier, the rate can be small enough that the transition is a rare event.
To illuminate these rare events, one can look at transition paths where the endpoints are pinned in different free energy basins and where the atomic motion is described by Brownian dynamics.
The understanding of these pinned Brownian paths is important for uncovering the driving mechanisms of molecular transitions, for example in the biochemistry of protein folding{\cite{proteinfolding}}.
Onsager and Machlup{\cite{Onsager:1953}} used the overdamped Langevin equation (Brownian Dynamics) as a basis for determining path probabilities in terms of the path variables themselves.
Their work gave rise to an expression that has become known as the Onsager-Machlup (OM) functional.
The particle dynamics is described by the Brownian stochastic differential equation (SDE) and is dependent on the choice of the time increment $\Delta t$.
Many{\cite{eric2004,facc2006,millerpred:2007,franklin2007minactionpath, PATH2016}} have used this approach to tackle various problems in path sampling.
The path sampling approach has the useful feature where the initial and final states can be defined to be in two separate free energy basins, and the rare transition paths between these basins can be efficiently explored.

The continuous-time limit  ($\Delta t \rightarrow 0$) of the OM functional is used to construct the Ito-Girsanov (IG) expression{\cite{Oksendal2003}}.
We examine computer algorithms that use the IG expression and that are designed to sample the aforementioned Brownian paths with fixed end points.
Although the Feynman-Kac formula has been proven for the infinite dimensional case of continuous time, its behavior when the dimensionality is finite ($\Delta t >0$) makes the limiting process only conditionally convergent since the kernal (of the path integral) approaches zero as the normalization factor goes to infinity in the $\Delta t  \rightarrow 0$ limit{\cite{lorinczi2011feynman}}.
In this paper we show that, when dimensionality is finite ($\Delta t >0$), the use of the  IG expression in path sampling algorithms does not have a firm mathematical foundation.
Here it is important to note that the Ito-Girsanov (IG) expression is not a measure but the change of measure between paths that correspond to two different SDEs or diffusions; one of which is free Brownian motion
Using this IG expression as a measure to compare the relative probabilities of two arbitrary paths in computational Metropolis  algorithms can, and does, lead to unphysical results as has been previously shown{\cite{malsom2015phd,PhysRevE.94.042131}}.
In addition, we emphasize that minimizers of the IG expression, used for example, in large computer codes{\cite{PATH2016}}, can be unphysical{\cite{MPP:2010}}.
When using a nonzero time increment, one has discarded an infinite number of degrees of freedom; the lack of the corresponding fluctuations is not sufficient to smear out the sharp minima that the IG expression contains.
We also examine the role of the (time) length of the path, and refute the unsubstantiated claim  made in the literature{\cite{chandrasekaran2017augmenting}} that short and long paths behave differently.
In particular, the value of the OM functional cannot be used to distinguish between short paths that do or do not cross an energy barrier.
Additionally, in the appendix of this work, we look at a particle moving in simple two-dimension potential with multiple (two) saddle points. 
In this example, the minimizer corresponds to a route through the wrong saddle point, again demonstrating the unreliability of the approach. 

For the sake of clarity, in the main body of this work,  we restrict our discussion to the problem of a single particle moving in one spatial dimension, $x$, under the influence of a potential, $U(x)$.
The Brownian SDE can be written as
\begin{equation}
dx_t=F(x_t)\,dt +\sqrt{2 \, \epsilon \,} \, dW_t
\label{SDE}
\end{equation}
where the position $x$ is a function of time $t$, $F(x) = -U'(x)$ is the conservative force, $\epsilon$ is the temperature, and the last term is the standard Wiener process which represent the random forces (white noise).
Computationally the Euler-Maruyama method{\cite{Maruyama1955} transforms the SDE for a nonzero time increment $\Delta t$ to the following, 
\begin{equation}
x_{n+1} =x_n  + F(x_n)\, \Delta t +\sqrt{2 \epsilon \, \Delta t \,}\, \xi_n
\label{SDEdt}
\end{equation}
where the index $n$ is related to the time $t=n \, \Delta t$, and $\xi_n$ is a Gaussian distributed random variable having mean zero and unit variance.
Onsager and Machlup{\cite{Onsager:1953}} then pointed out the path probability, $\mathbb{P}_{OM}$, was given by the product  
\begin{equation}   \mathbb{P}_{OM}\propto \Pi_n \exp{\! \big( - \xi_n^2/2 \big)} \label{OMp} \end{equation}
and  with Equation {\ref{SDEdt}}, could be expressed as $\mathbb{P}_{OM} \propto \exp{\!\big(-\mathbb{I}_{OM} \big)}$ in which $\mathbb{I}_{OM}$ is a function of the path variables, that is,
\begin{equation}
\mathbb{I}_{OM} = \frac{\Delta t}{4 \, \epsilon} \sum_n \Big|   \frac{x_{n+1}-x_n}  {\Delta t} \, - \, F(x_n) \,  \Big|^2 \ .
\label{defOM}
\end{equation}
This expression for $\mathbb{I}_{OM}$ defines the OM functional  away from the $\Delta t \rightarrow 0$ limit.

\begin{figure}[b]
\includegraphics[width=.9 \columnwidth]{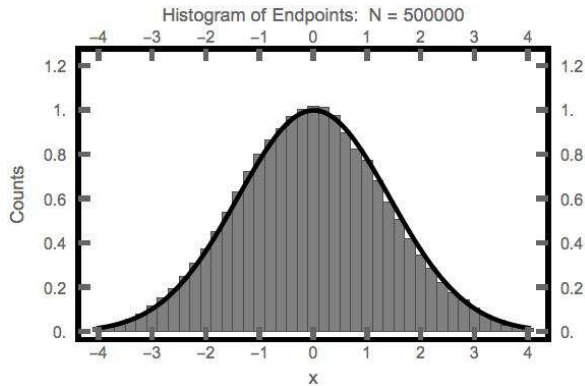}
\caption{Histogram of endpoints plotted along with the expected result (solid line).
Calculation uses $\epsilon =1 $, $T=1$, and $\Delta t = 2 \, \times \,10^{-6}$.}
\label{hist}
\end{figure}

Much later Graham{\cite{graham1977path}} and others{\cite{bach1977functionals,durr1978onsager,hunt1981path,dai1989rigorous}} indicated that the OM functional could be interpreted as a "thermodynamic action" and used in a similar manner that is done in the path integral formulation of quantum mechanics{\cite{mckane2009stochastic,orland2011generating}}.
That is, the OM functional could be used to produce a thermodynamically relevant ensemble of paths with fixed starting and ending points: the starting point is $x(0)=x_-$ and the endpoint is fixed at $x(T)=x_+$.
We define $P_{\Delta t}(x_+,\,T \,| \, x_-,\,0)$, an analog to the single particle propagator from quantum mechanics, to be the probability that the particle starts at $t=0$ with position $x_-$ and at a time $t=T$ the particle position is $x_+$.
This function can be expressed as
\begin{equation}
P_{\Delta t}(x_+,\,T \,| \, x_-,\,0) \propto \int \mathcal{D}[x]  \, \exp{\!\big(-\mathbb{I}_{OM}   \big) }
\label{PIOM}
\end{equation}
with the proviso that the endpoints are fixed as specified above.

In the large $T$ limit, to be consistent with thermodynamics, this probability is independent of $x_-$ and is given in terms of the Boltzmann factor as
\begin{equation}
\lim_{T \to \infty } P_{\Delta t}(x_+,\,T \,| \, x_-,\,0) = \frac{\exp{ \!\big( -U(x_+) /\epsilon \big)  }}{Z}
\label{finiteProb}
\end{equation}
where $Z$ is the normalization factor (the partition function).
Note that when the force is "globally Lipschitz," the limit $\Delta t \rightarrow 0$ is guaranteed to converge to the correct thermodynamic value{\cite{Mackevi2003}}; otherwise it may not.

\begin{figure}[b]
\includegraphics[width=.95 \columnwidth]{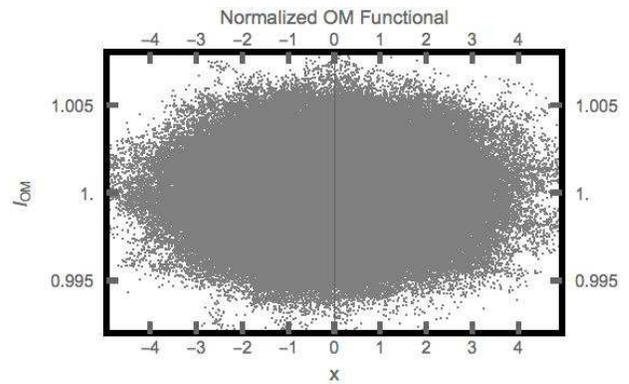}
\caption{The normalized value of $\mathbb{I}_{OM} $ plotted as a function of ending position for the zero potential case.
The normalization factor is simply $.5 \, T/ \Delta t$, where $T$ is the duration of the path.}
\label{GOM}
\end{figure}
 
Before moving to the continuous-time measure, let us examine values of the OM functionals that is consistent with Equation \ref{SDEdt}.
In the case where the potential is zero, the conditional probability  in Equation \ref{finiteProb} can be calculated and is given by $P_{\Delta t}(x_+,\,T \,| \, x_-,\,0)  =  (4 \, \pi \,\epsilon \, T)^{-1/2} \exp{\! \Big(-  \frac{  (x_+ - x_-)^2}   {4\, \epsilon \, T}  \Big)}$.
To understand how the OM functional behaves in this special case with zero potential, we iterate Equation {\ref{SDEdt}} many times (500,000) with $x_-=0$ and examine the ending points of the trajectories.
In Figure ${\ref{hist}}$ we show a histogram of the ending points of these iterations.
It is not surprising that this simulation yields the correct Gaussian distribution of endpoints; the sums of Gaussian deviates are Gaussian distributed.
We also looked at the value of $\mathbb{I}_{OM} $ and how it varies as a function of the ending position.
Examining how the value of $\mathbb{I}_{OM}$ varies as a function of these endpoints yields another unsurprising distribution.
In Figure {\ref{GOM}}, we see that the value of $\mathbb{I}_{OM} $  is  almost \textit{flat}, with at most a $1\%$ variation.
Note the absence of any correlation between the value of $\mathbb{I}_{OM} $ and the value of the endpoint (of the trajectory).
This is what is expected since the variance of each set of deviates is independent of the sum of the deviates.

We now repeat the above exercise for the case where the potential is nonzero.
In particular, we chose a potential that has been studied before{\cite{PhysRevE.94.042131}}, namely 
\begin{equation}
U(x) = \Big( \, \frac 5 2 \, x+1 \Big)^2 \  \Big( \, \frac 5 8  \, x-1 \Big)^8  
\label{upot}
\end{equation}
which has two degenerate wells, a unit energy barrier at the origin, a broad well on the right and a narrow (quadratic) well on the left (of the origin).
By again iterating Equation 2 many times (500,000) with $x_-=-0.4$ and examining the distribution of endpoints (see Figure 3), we find the ending points are very represenaitive of the Boltzmann distribution.
Note that the information about the starting position was lost.
In Figure 4, we again show that the value of $\mathbb{I}_{OM} $  is again essentially flat, even when the potential is nonzero.
As shown in this figure, no correlation exists between the value of $\mathbb{I}_{OM} $ and the ending point of the path.
As previously stated:{\cite{PhysRevE.94.042131}} the noise originates in the heat reservoir which has no knowledge of what system is under investigation.

\begin{figure}[t]
\includegraphics[width=.9 \columnwidth]{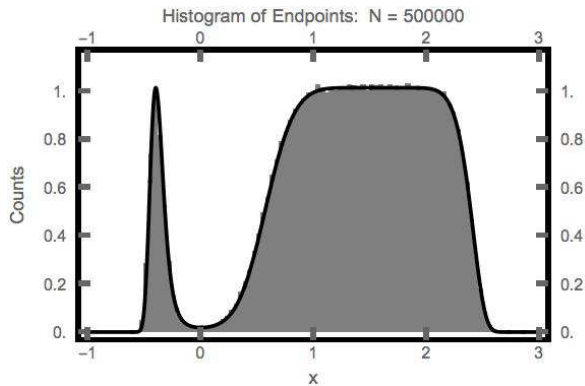}
\caption{Histogram of endpoints plotted along with the Boltzmann factor (solid line) using the potential given in Equation {\ref{upot}}.
Calculation uses $\epsilon =0.25 $, $T=500$, and $\Delta t = 1 \, \times \,10^{-3}$.}
\label{histfs}
\end{figure}

\begin{figure}[b]
\includegraphics[width=.95 \columnwidth]{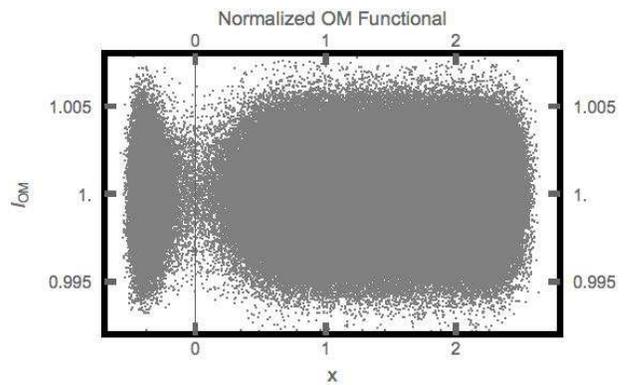}
\caption{For long paths, the normalized value of $\mathbb{I}_{OM} $ plotted as a function of endpoint for the potential of Equation {\ref{upot}}.
The normalization factor is simply $.5 \,  T/ \Delta t$, where $T=500$ is the duration of the path.}
\label{fsOM}
\end{figure}

\begin{figure}[b]
\includegraphics[width=.95 \columnwidth]{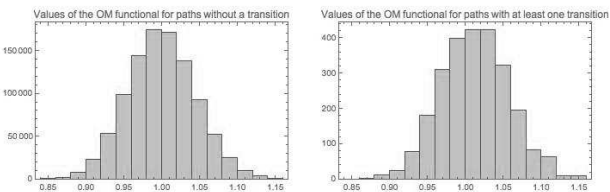}
\caption{For short paths, the histogram of the number of occurrences of the normalized value of $\mathbb{I}_{OM} $  for the potential of Equation {\ref{upot}}.
The normalization factor is simply $.5 \,  T/ \Delta t$, where $T=1$ is the duration of the path, and $\Delta t=10^{-3}$. 
The histogram on the left (right) plots the OM functional for the paths that made no (at least one) transition.}
\label{fsOMshort}
\end{figure}

Since $\mathbb{I}_{OM} $  is flat, the distribution of endpoints is simply governed by the path multiplicity.
This multiplicity can be viewed as the inverse Jacobian of the transformation that relates the probability of paths to the single particle propagator, $P_{\Delta t}$, Equation {\ref{finiteProb}}.
In the case where the potential is zero, the number of paths with the Gaussian deviates summing to a large number is small compared to the number of paths where the sum is small.
When the potential is nonzero, for long enough paths, positions are being visited in a manner that is consistent with the Boltzmann distribution.
The path is more likely to visit points with low values of the potential compared to those with high values.
The position at time $T$ is not special, and thus the endpoints are also distributed according to the Boltzmann probability (as long as $T$ is significantly larger than the barrier crossing time).

We now address what happens when the path length is shorter  than the inverse of the transition rate.
We repeated the above calculation only changing the path length, decreasing $T$ from $500$ to $1$.
As expected we find that only a small fraction (0.0026) of the $10^6$ trajectories contained a barrier hopping event.
The paths had the same starting point $x_-= -0.4$. 
If a position along the path exceeded $1.6$, we classified the path as containing a transition. 
We tabulated the values of the OM function for all $10^6$ paths, and separately plotted (in Figure {\ref{fsOMshort})} the histogram for paths that contained a transition and a second for those that did not.  
From Figure {\ref{fsOMshort}}, one sees that the values of the OM functional cannot be used to differentiate the two types of paths.
The rareness of the transition does not come from the value of the OM functional; the OM functional is flat: it is the same for all paths.
It is simply an uncommon event to have the noise sequence that is sufficiently coherent to propel the particle over a barrier.
It is much more common for the particle to fail to cross over to the other well.
In Figure {\ref{fsOMshort}} we show that short and long paths behave in the same way: the path multiplicity governs the position at the end of the path and not the value of the OM functional. 
Unlike for long paths, the end point for short paths is not govern by the Boltzmann distribution, but by the height of the energy barriers.

Now we turn to the question of what happens in the continuous-time limit.
The approach has been to first take the continuous-time limit of the OM functional{\cite{DurrBach:1978,Beskos2008}}.
In particular, the limit can be extracted{\cite{orland2011generating}} from the Fokker-Planck equation and written, in an informal manner, as
\begin{equation}
  \begin{split}
\mathbb{I}_{IG}  = \lim_{\Delta t \to 0} \  &  \mathbb{I}_{OM}   
 =  \frac{U(x_+)-U(x_-)}{2 \, \epsilon}   \qquad   \qquad  \\  &+   \frac{1}{2 \, \epsilon}  \int_0^T \!\! dt \, \Bigg( \frac{1}{2} \Big(\frac{dx}{dt} \Big)^2 +G(x_t)\Bigg)
  \end{split}
  \label{Iig}
\end{equation}
with $G(x) = \frac{1}{2} |\,F(x)\,|^2 - \epsilon \, U''(x)$.
The subscript $IG$ is used because one gets a similar form using the Radon-Nikodym derivative, Girsanov's theorem and Ito's lemma{\cite{Oksendal2003}}.
At this point, it is instructive to again point out that, away from the continuous-time limit, the value of $I_{OM}$ is effectively flat, and continues to be so, even as $\Delta t$ gets smaller and smaller.
Note that  when $\Delta t$ is small (but nonzero) $I_{OM}$ is proportional to the variance of the Gaussian random numbers, see Equations $\ref{OMp}$ and $\ref{defOM}$, for any path that is compatible with the Wiener process, i.e. thermodynamics.

For any $\Delta t >0$, no path is special in that the value of $\mathbb{I}_{OM}$  is essentially the same for all paths;
no path is anointed to be the "most probable."
In the continuous-time limit, the infinite number of degrees of freedom wash out the structure in the IG expression.
But, when working with computer algorithms, one only has a finite number of degrees of freedom and those are not sufficient to produce the fluctuations needed to smear out the structure.

Further examination of Equation ${\ref{Iig}}$ leads to a more serious  problem: $\mathbb{I}_{IG} $ diverges in the continuous-time limit (as $\Delta t\rightarrow 0 $).
The quantity $\sum_n \,  \Delta t \,  \Big(\Delta x / \Delta t \Big)^2$ is proportional to the ratio $T/ \Delta t$ which diverges in the continuous-time limit.
We now must examine the ramifications of this divergence which is dire for the classical problem studied here but, as discussed below, of little consequence when it occurs when using quantum Feynman path integrals..

It is illuminating to compare the discrete action ($\mathbb{I}_{OM}$) and the continuous-time limit ($\mathbb{I}_{IG}$) for the case of a vanishing force.
Without the force, the integral in Equation \ref{PIOM} can be evaluated by factoring $\mathbb{I}_{OM}$, namely 
\begin{equation}
  \begin{split}
P_{\Delta t}(x_+,T & | x_-,0)   \propto      \Pi_n   \Bigg( \! \int dx_n \, \exp{\Bigg\{   - \frac{  \big(x_{n+1}-x_n \big)^2}  { 4 \,
 \epsilon \, \Delta t}   \Bigg\} \Bigg)}   \\ 
& =  (4 \, \pi \,\epsilon \, T)^{-1/2}
\exp{\! \Bigg\{  -  \frac{  (x_+ - x_-)^2}   {4\, \epsilon \, T}  \Bigg\}  }    \end{split}.
\label{prop}
\end{equation}
Evaluating the path probability in this way gives a result which is consistent with free diffusion.
This evaluation uses the same regularization that is used for quantum path integrals{\cite{Shankar}}.
However, unlike the quantum case, the limit is not absolutely convergent, even using the regularization trick.
If one uses $\mathbb{I}_{IG}$, the continuous-time limit of  $\mathbb{I}_{OM}$, a different result is found for this simple case:  
$\lim_{\Delta t \rightarrow 0} P_{\Delta t}$ becomes an ill-defined product of $0$ and $\infty$; the latter coming from the normalization inherent in Equation $\ref{PIOM}$.
Clearly, even in this very simple case, these two limiting procedures are not compatible.

Additionally, since the divergence of the  $\mathbb{I}_{IG}$ is independent of the potential, this limit of $P_{\Delta t}$  is ill-defined even when the potential is nonzero.
Thus $P_{\Delta t}$  in the continuous-time limit is only conditionally convergent.
The Fokker-Planck equation describes the time evolution of the spread of the probability density function.
And the Feynman-Kac Theorem connects the Fokker-Planck equation to the Brownian dynamics in continuous time.
Note that although the Feynman-Kac Theorem holds in continuous time,  the zero coming from $\exp{\! \big( -\mathbb{I}_{IG} \big) }$ and the infinity coming from the normalization combine to get something sensible{\cite{lorinczi2011feynman}} only within the associated measure theory.
However computational methods must use a finite representation of path, and thus do not contain the infinite (path) multiplicity needed to cancel the vanishing of  $\exp{\! \big( -\mathbb{I}_{IG} \big) }$ as an infinite number of degrees of freedom have been ignored.
Thus $\mathbb{I}_{IG}$ does not possess the mathematical foundation that it needs to be considered a path probability measure in finite dimensions, that is, when $\Delta t >0$.

In this way, classical path integrals differ from their quantum mechanical counterparts.
The factor of $i = \sqrt{-1}$ in front of the action alters the convergence property of the free quantum propagator (as compared to the classical, Brownian case).
In the classical case, all contributions to $P_{\Delta t}$ are nonnegative.
For the quantum propagator, the factor of $i = \sqrt{-1}$ turns the role of $\mathbb{I}_{OM}$ into a simple phase, and  the relevant  term is its modulo $2\,\pi$.
Thus the main contributions{\cite{feynman1965quantum}} to quantum single particle propagator come from the Feynman paths where this phase factor is "stationary."  
Contributions to the quantum propagator are then of both signs{\cite{dirac1933}}.
Thus in the quantum case, the limiting process does not depend on the order of operations.
Computational algorithms that produce accurate results when $\Delta t >0$ in the quantum case, fail in the classical case of  evaluating the probability of pinned Brownian paths when using the continuous-time limit of the OM functional. 

The observations reported here have an impact on other areas of research.
For example, in the numerical methods designed by Kappen and Ruiz{\cite{kappen2016adaptive}} to analyze path integral control problems in continuous time, the sampling of the uncontrolled dynamics is used.
If the dynamics is doubly constrained, this numerical method will suffer from the same problems as described above.
Other works{\cite{hartmann2012efficient, opper2017estimator}} are derived in the continuous-time limit,  and thus those algorithms may produce unphysical results for the same reason.
Path integral methods for these problems originate from a limit that is only conditionally convergent.
The derived computer algorithms  do not possess the mathematical foundation to ensure that the ensuing calculations remain physical.
Moreover this paper may have ramifications for techniques used for data assimilation in continuous time as some are "in part adopted from the theory of optimal nonlinear control."{\cite{brocker2010variational}}

In summary, while the continuous-time limit of the OM functional is given by the Ito-Girsanov (IG) change of measure,
 using this expression in an algorithm to sample transition paths can produce grossly unphysical results{\cite{malsom2015phd,PhysRevE.94.042131}}.
Here, we have shown that the reason behind this result is the absence of a firm mathematical foundation.
The divergence of the continuous-time limit of the OM functional renders the IG expression useless if the goal is to develop computer algorithms for constructing a physically relevant, thermal distribution of pinned paths.
This divergence cannot be offset without the infinite multiplicity of paths that only exist in continuous time.
Thus computer algorithms that use discrete time lack this infinite multiplicity even as the time increment becomes tiny.
It is in this sense that the continuous-time limit can be considered singular.
The Feynman-Kac theorem holds only in the infinite dimensional space of continuous time and not in the finite dimensional path space where computational algorithms operate.
The computational algorithms used to understand quantum path integrals fail when used for pinned Brownian paths.
Thus the use of  IG expression should be avoided.
If one insists on using the  IG expression{\cite{franklin2007minactionpath,eric2004,PATH2016,fujisaki2010,orland2015,carter2017combining,chandrasekaran2017augmenting}}, one must be prepared to find unphysical results.
A variety of examples have been published before{\cite{MPP:2010} demonstrating the unphysical behavior of minimizers of the continuous-time limit of the OM functional.
In the appendix below, we explore yet another example where the minimizer is obviously unphysical in that it identifies a transition state that is inconsistent with thermodynamics.
In a recent paper{\cite{PhysRevE.94.042131}}, we showed that even if one used the Ito-Girsanov expression in robust sampling algorithms, one may obtain unphysical results.
Taken together with the above, these illustrate the unreliability of interpreting the continuous-time limit of the OM functional as a probability measure. 
Here we trace the origins of this behavior back to the limiting process, which is only conditionally convergent.
Similar problems exist in a wide range of areas that use doubly-constrained Brownian paths in continuous time and thus have the same ill-defined mathematical foundation.
However, the limiting process of quantum counterpart to this classical path representation does not have such a problematic limit. 
The factor of $i=\sqrt{-1}$ turns the OM functional into a phase, eliminating the divergence of the single particle propagator in the continuous-time limit.

\appendix 
\setcounter{figure}{0} \renewcommand{\thefigure}{A.\arabic{figure}}
\setcounter{equation}{0} \renewcommand{\theequation}{A.\arabic{equation}}
\section*{Appendix}
\begin{figure}[t]
\includegraphics[width=.95 \columnwidth]{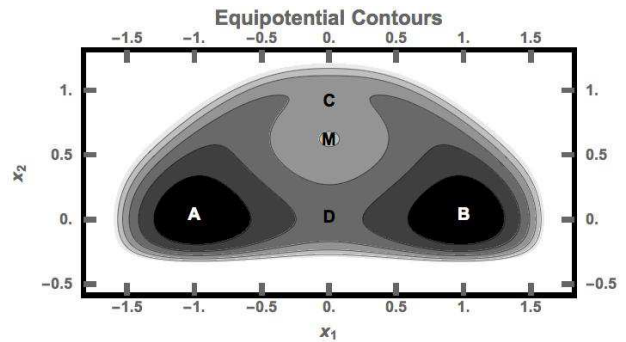}
\caption{Contour plot of the potential described in Equation $\ref{eq:V2d}$.
The wells of equal depth are denoted by $A$ and $B$.
The saddle point in the narrow channel is denote by $C$ and the one in the wide channel by $D$.
The local maximum is denote by $M$.}
\label{pot2d}
\end{figure}

It is important to note that some researchers have already incorporated the minimization of the OM functional (the thermodynamic action) into large codes{\cite{PATH2016,fujisaki2010,carter2017combining,chandrasekaran2017augmenting}}.
Researchers who dogmatically use the MinActionPath{\cite{franklin2007minactionpath}} or PATH{\cite{PATH2016}} codes should recognize that this procedure is unreliable in that their results may be extremely unphysical.
We provide the following example to illustrate one of the many ways where one can go astray: a case where the minimizer of the OM function identifies a transition state and pathway that conflicts with thermodynamics.
Consider a particle moving in the following two-dimensional potential:  
\begin{align}
U(x_1,x_2)=&   4 (x_1^2 + x_2^2 - 1)^2 x_2^2 -  
\exp\bigl(-4( (x_1 - 1)^2 + x_2^2 )\bigr)\notag\\ 
& - \exp\bigl(-4( (x_1 + 1)^2 + x_2^2 )\bigr)  +  \alpha_1 \,  x_2 
\notag\\
&+\exp\bigl(8(x_1 - 1.5)\bigr) +  \exp\bigl(-8(x_1 + 1.5)\bigr)\notag\\
&+ \exp\bigl(  \alpha_2 (x_2 + 0.25)\bigr) + 0.2\exp\bigl(- 8 x_1^2\bigr).
\label{eq:V2d}
\end{align}
The choice $\alpha_1=0$ and $\alpha_2 = -12.163839304988416 $ was investigated before.{\cite{MPP:2010}}
Here we keep the same value of $\alpha_2$ and choose $\alpha_1 = 1/2$.
This potential looks fairly innocuous, see Figure $\ref{pot2d}$; it possesses two (degenerate) minima, one local maximum and two saddle points.
Two different channels exist for the particle to move from well $A$ to well $B$.
The barrier at $C$ is about $42\%$ larger than the barrier at $D$.
The direct path lies in a wide channel and has the smaller energy barrier at $D$.
The circular path lies in a narrow channel and has the larger energy barrier at $C$.

\begin{figure}[t]
\includegraphics[width=.95 \columnwidth]{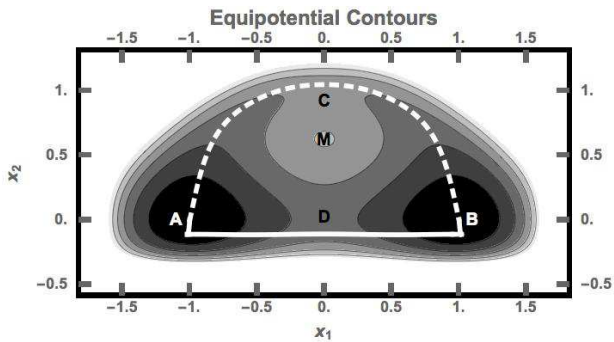}
\caption{The paths that generate local minima of the OM functional.
The solid line is the direct route, passes through the lowest barrier at $D$.
The dotted line is the circular route, passes through the higher of the two barriers but gives the lowest value for the OM functional.}
\label{pot2dp}
\end{figure}

We then look for minimizers of the continuous-time limit of the OM functional when the paths are constrained to start at $A$ and end at $B$.
We chose a temperature to be one-quarter of the barrier height at $D$:  $\epsilon = (U_D-U_A)/4$.
We studied two different path lengths, $T=5$ and $T=25$.
Using forward integration of the SDE (2-dimensional generalization of Equation $\ref{SDEdt}$), we found the transition rate, $R$, is about $R \approx 0.02$ which gives an average time between transitions of about $50$.
The smaller choice for $T$ is roughly $1/10$ of the inverse transition rate.
Thermodynamics would tell us that the particle is more likely to pass over the lower energy barrier (at $D$), and this is confirmed by forward integration of the SDE.
The minimizers of the IG expression are shown in Figure $\ref{pot2dp}$; the dependence of $T$ cannot be seen at the level of detail displayed in the figure.
For both values of $T$, we obtain the unphysical result that the minimum value for the continuous-time limit of the OM functional is attained by the path that passes over the larger barrier (at $C$).
This result is consistent with the results discussed in the main section of this paper; long and short paths behave similarly.  
Furthermore this minimizer indicates the particle tends to spend time near the points with $x_1=  \pm 0.266$ and $x_2 =1.014$ which do not correspond to any feature of the potential $U$, defined in Equation $\ref{eq:V2d}$, but are artifacts of including the Laplacian in the definition of $G$.
For values of $T \ge 25$, the minimizing paths would be very similar; for these larger values of $T$, the particle simply spends a longer time at the same unphysical positions, i.e. the minima of $G$.{\cite{MPP:2010}}
As $T$ becomes smaller than $5$, eventually the minimizer would become ballistic in that the values of the potential becomes irrelevant.

Fluctuations play an important role by washing out sharp structure in the IG expression and thus diminishing the importance of the minimizer. 
Note that when $\Delta t >0$, an infinite number of degrees of freedom are discarded and the remaining (finite in number) ones may not be sufficient to do this{\cite{PhysRevE.94.042131}}. 
For computer algorithms, one is relegated to using a finite representation, and thus unphysical effects creep into the resulting calculations.
With  this example, we aim to remind researchers{\cite{chandrasekaran2017augmenting,orland2015}} that minimizers of the IG expression, the continuous-time limit of the OM functional, are unreliable in that they can produce unphysical results including identifying unphysical transition states and pathways.
The unphysical results found in the above example are quite insensitive to the path length in contrast to the speculative remark in the literatures{\cite{chandrasekaran2017augmenting}}.

\section*{Acknowledgments}
 We wish to thank Carsten Hartmann, Gideon Simpson, Robin Ball, Andrew Stuart, and Henri Orland for many discussions.

\bibliographystyle{apsrev4-1}

\end{document}